\documentclass[12pt]{article}
\usepackage{longtable,supertabular}
\usepackage{amsmath}
\usepackage{amssymb}
\usepackage{latexsym}
\usepackage{cite}

\title{\bf Covering Codes, List-Decodable Codes and Strong Singleton-Like Bounds in the Sum-Rank Metric  }
\author{Hao Chen
  \thanks{Hao Chen is with the College of Information Science and Technology/Cyber Security, Jinan University, Guangzhou, Guangdong Province, 510632, China, haochen@jnu.edu.cn. This research was supported by NSFC Grant 62032009.}}

\begin{document}

\maketitle
\begin{abstract}
Codes in the sum-rank metric have received many attentions in recent years, since they have wide applications in the multishot network coding, the space-time coding and the distributed storage. Fundamental bounds, some explicit or probabilistic constructions of sum-rank codes and their decoding algorithms have been developed in previous papers. In this paper, we construct covering codes in the sum-rank metric from covering codes in the Hamming metric. Then some upper bounds on sizes of covering codes in the sum-rank metric are presented. Block length functions of covering codes in the sum-rank metric are also introduced and studied. As applications of our upper bounds on covering codes in the sum-rank metric and block length functions, several strong Singleton-like bounds on sum-rank codes are proposed and proved. These strong Singleton-like bounds are much stronger than the Singleton-like bound for sum-rank codes, when block lengths are larger. An upper bound on sizes of list-decodable codes in the sum-rank metric is also given, which leads to an asymptotic bound on list-decodability of sum-rank codes. We also give upper bounds on block lengths of general binary MSRD codes. \\

\end{abstract}

\section{Introduction}

\subsection{Codes in the Hamming metric and the rank-metric}

Let us recall basic facts of error-correcting codes in the Hamming metric and the rank-metric. For a vector ${\bf a}=(a_1, \ldots, a_n) \in {\bf F}_q^n$, the Hamming weight $wt_H({\bf a})$ of ${\bf a}$ is the cardinality of its support $$supp({\bf a})=\{i: a_i\neq 0\}.$$ The Hamming distance $d_H({\bf a}, {\bf b})$ between two vectors ${\bf a}$ and ${\bf b}$ is defined to be $wt_H({\bf a}-{\bf b})$. For a code ${\bf C} \subset {\bf F}_q^n$ of dimension $k$, its Hamming distance $$d_H=\min_{{\bf a} \neq {\bf b}} \{d_H({\bf a}, {\bf b}): {\bf a}, {\bf b} \in {\bf C}\},$$  is the minimum of Hamming distances $d_H({\bf a}, {\bf b})$ between any two different codewords ${\bf a}$ and ${\bf b}$ in ${\bf C}$. It is well-known that the Hamming distance of a linear code ${\bf C}$ is the minimum Hamming weight of its non-zero codewords. We refer to \cite{HP} for the theory of error-correcting codes in the Hamming metric.   For a linear $[n, k, d_H]_q$ code, the Singleton bound asserts $d_H \leq n-k+1$, see \cite{Singleton}. When the equality holds, this code is called an maximal distance separable (MDS) code. Reed-Solomon codes are the most well-known MDS codes, which were constructed in 1960, see \cite{RS} and \cite[Chapter 5]{HP}. The study of MDS codes has been one of the central topic within coding theory.\\

The rank-metric on the space ${\bf F}_q^{(n, n)}$ of size $n \times n$ matrices over ${\bf F}_q$ is defined by the ranks of matrices, $d_r(A,B)= rank(A-B)$. The minimum rank-distance of a code ${\bf C} \subset {\bf F}_q^{(m, n)}$  is $$d_r({\bf C})=\min_{A\neq B} \{d_r(A,B): A, B \in {\bf C} \}$$  For a code ${\bf C}$ in ${\bf F}_q^{(m, n)}$ with the minimum rank distance $d_r({\bf C}) \geq d$, the Singleton bound in the rank-metric asserts that the number of codewords in ${\bf C}$ is upper bounded by $q^{n(n-d+1)}$ , see \cite{Gabidulin}.  A code attaining this bound is called a maximal rank distance (MRD) code. The Gabidulin code consisting of ${\bf F}_q$ linear mappings on ${\bf F}_q^n \cong {\bf F}_{q^n}$ defined by $q$-polynomials $a_0x+a_1x^q+\cdots+a_ix^{q^i}+\cdots+a_tx^{q^t}$, where $a_t,\ldots,a_0 \in {\bf F}_{q^n}$ are arbitrary elements in ${\bf F}_{q^n}$, see \cite{Gabidulin}. The rank-distance of the Gabidulin code is at least $n-t$ since there are at most $q^t$ roots in ${\bf F}_{q^n}$ for each such $q$-polynomial. There are  $q^{n(t+1)}$ such $q$-polynomials. Hence the size of the Gabidulin code is $q^{n(t+1)}$ and it is an MRD code.  \\

For a code ${\bf C}$ in the Hamming metric space ${\bf F}_q^n$, we define its covering radius by $$R({\bf C})=\max_{{\bf x} \in {\bf F}_q^n} \min_{{\bf c} \in {\bf C}} \{wt({\bf x}-{\bf c})\}.$$ Hence the Hamming balls $$B({\bf x}, R({\bf C}))=\{{\bf y} \in {\bf F}_q^n: wt_H({\bf y}-{\bf x}) \leq R_H({\bf C})\}$$ centered at all codewords $x \in {\bf C}$, with the radius $R({\bf  C})$ cover the whole Hamming metric space ${\bf F}_q^n$, and moreover this radius is the smallest possible such radius. The basic goal of covering problem is as follow. Given a positive integer $R<n$, to find a small code in the Hamming metric space ${\bf F}_q^n$ such that its covering radius is at most $R$. Actually the covering radius $R({\bf C})$ of a linear $[n, k]_q$ code ${\bf C} \subset {\bf F}_q^n$ can be determined as follows. If $H$ is any $(n-k) \times n$ parity check matrix of ${\bf C}$, $R({\bf C})$ is the least integer such that every vector in ${\bf F}_q^{n-k}$ can be represented as ${\bf F}_q$ linear combinations of $R({\bf C})$ or fewer columns of $H$, see \cite{CHLL}. It follows the redundancy upper bound $$R({\bf C}) \leq n-k$$ for a linear $[n, k]_q$ code, see \cite{CHLL}. The covering radius of the Reed-Solomon $[n, k, n-k+1]_q$ code is exactly $n-k$, see \cite{CHLL}. From the above interpretation of the covering radius from the parity check matrix, it is obvious that if ${\bf C}$ is a covering code in ${\bf F}_q^n$ with the radius $R$, then ${\bf C} \otimes {\bf F}_{q^h} \subset {\bf F}_{q^h}$ is a covering code with the radius at most $hR$.\\

Let $n$ be a fixed positive integer and $q$ be a fixed prime power, for a given positive integer $R<n$, we denote $K_q(n, R)$ the minimal size of a code ${\bf C} \subset {\bf F}_q^n$ with the covering radius smaller than or equal to $R$. Set $ \frac{log_q K_q(n, \rho n)}{n}=k_n(q, \rho)$. The following asymptotic bound is well-known,  $$1-H_q(\rho) \leq k_n(q,\rho) \leq 1-H_q(\rho)+O(\frac{logn}{n}),$$ where $H_q(r)=rlog_q(q-1)-rlog_q r-(1-r)log_q (1-r)$ is the $q$-ary entropy function, see \cite[Chapter 12]{CHLL}. Covering radii of Reed-Muller codes, BCH codes in the Hamming metric have been studied for many years, we refer to \cite{CHLL} and \cite{Le} for results on covering radii of BCH codes and Reed-Muller codes.\\

Let $l_q(r,R)$ be the smallest length $t$ such that there is a linear covering code ${\bf C} \subset {\bf F}_q^t$ with the covering radius $R$ and the redundancy at least $r$. This parameter is called the length function of covering codes, which was introduced in \cite{BPW} in 1989, see also \cite{LP} and \cite[Section 11.5]{HP}. For upper bounds on some length functions, we refer to \cite{DMP,DMP2}. In this paper we introduce similar block length functions for covering codes in the sum-rank metric.\\

\subsection{Sum-rank codes}

Codes in the sum-rank metric have applications in multishot network coding, see \cite{MK19,NPS,NU}, space-time coding, see \cite{SK}, and coding for distributed storage see \cite{CMST,MK,MP1}. For fundamental properties and constructions of sum-rank codes, we refer to \cite{MK,MP1,BGR,BGR1,CGLGMP,OPB,MP21,MP22}. Now we recall some basic concepts and results for sum-rank codes in \cite{BGR}. Let $n_i \leq m_i$ be $2t$ positive integers satisfying $m_1 \geq m_2 \cdots \geq m_t$. Set $N=n_1+\cdots+n_t$, ${\bf n}=(n_1, \ldots, n_t)$ and ${\bf m}=(m_1, \ldots, m_t)$.  Let $${\bf F}_q^{(n_1, m_1), \ldots,(n_t, m_t)}={\bf F}_q^{n_1 \times m_1} \bigoplus \cdots \bigoplus {\bf F}_q^{n_t \times m_t}$$ be the set of all ${\bf x}=({\bf x}_1,\ldots,{\bf x}_t)$, where ${\bf x}_i \in {\bf F}_q^{n_i \times m_i}$, $i=1,\ldots,t$, is the $n_i \times m_i$ matrix over ${\bf F}_q$. We call $n_i \times m_i$, $i=1, \ldots, t$,  matrix sizes of sum-rank-metric codes. Set $wt_{sr}({\bf x}_1, \ldots, {\bf x}_t)=rank({\bf x}_1)+\cdots+rank({\bf x}_t)$ and $$d_{sr}({\bf x},{\bf y})=wt_{sr}({\bf x}-{\bf y}),$$ for ${\bf x}, {\bf y} \in {\bf F}_q^{(n_1,m_1), \ldots,(n_t,m_t)}$. This is indeed a metric on ${\bf F}_q^{(n_1,m_1), \ldots,(n_t,m_t)}$.  The ball in the sum-rank-metric is defined by $$B_{sr}({\bf x}, d)=\{{\bf y} \in {\bf F}_q^{(n_1,m_1), \ldots, (n_t, m_t)}: wt_{sr}({\bf y}-{\bf x}) \leq d\}.$$

{\bf Definition 1.1.} {\em A sum-rank code ${\bf C} \subset {\bf F}_q^{(n_1,m_1), \ldots,(n_t,m_t)}$ is a subset in the finite metric space ${\bf F}_q^{(n_1,m_1), \ldots,(n_t,m_t)}$. Its minimum sum-rank distance is defined by $$d_{sr}({\bf C})=\min_{{\bf x} \neq {\bf y}, {\bf x}, {\bf y} \in {\bf C}} d_{sr}({\bf x}, {\bf y}).$$  The rate of this code is $R_{sr}=\frac{log_q |{\bf C}|}{\Sigma_{i=1}^t n_im_i}$. The relative distance is $\delta_{sr}=\frac{d_{sr}}{N}$.}\\

The following several special cases of parameters are important. When $t=1$, this is the rank-metric code case. When $n=1$, this is the Hamming error-correcting code case. Hence the sum-rank metric is a generalization and combination of the Hamming metric and the rank-metric. The basic goal of coding in the finite metric space ${\bf F}_q^{(n_1,m_1), \ldots, (n_t,m_t)}$ endowed with the sum-rank metric is to construct sum-rank codes with large sizes and large minimum sum-rank distances. For some basic upper bounds we refer to \cite[Section III \& IV]{BGR}. We denote the maximal possible size of sum-rank codes over ${\bf F}_q$ with the minimum sum-rank distance $d$ by ${\bf A}_q({\bf n}, {\bf m}, d)$ as in \cite{AKR}.\\

The Singleton-like bound for the sum-rank codes was proved in \cite{MK,BGR}. The general form Theorem III.2 in \cite{BGR} is as follows. Let the minimum sum-rank distance $d$ can be written uniquely as the form $d_{sr}=\Sigma_{i=1}^{j-1} n_i+\delta+1$ where $0 \leq \delta \leq n_j-1$, then $$|{\bf C}| \leq q^{\Sigma_{i=j}^t n_im_i-m_j\delta}.$$ The code attaining this bound is called maximal sum-rank distance (MSRD) code. When $m_1=\cdots=m_t=m$ this bound is of the form $$|{\bf C}| \leq q^{m(N-d_{sr}+1)}.$$  When $t=1$, it degenerates to the Singleton bound in the rank-metric. When $n_i=m_i=1$, for $i=1, \ldots, t$, it degenerates to the original Singleton bound for the Hamming metric codes, see \cite{Singleton}.\\

In the case of matrix size $1 \times 1$, MSRD codes are MDS codes in the Hamming metric. They have attracted a lot of attentions within the theory of error-correcting codes in the Hamming metric. For general matrix sizes, it is important to construct MSRD codes and to find various restrictions on parameters of MSRD codes in the sum-rank metric coding theory. When $t \leq q-1$ and $N \leq (q-1)m$, MSRD codes were constructed in \cite{MP1,MP22}. They are called linearized Reed-Solomon codes, we also refer to \cite{Neri} for the construction of twisted linearized Reed-Solomon codes.  When $t=q$, it was proved in \cite[Example VI.9]{BGR}, MSRD codes may not exist for some minimum sum-rank distance. In several other cases, for example, when the sum-metric minimum distance is $2$ or $N$, MSRD codes exist for all parameters, they were constructed in  \cite[Section VII]{BGR}. In \cite{MP20} more linear MSRD codes with the same matrix size (not square matrix) defined over smaller fields were constructed by extended Moore matrices. In our previous paper \cite{Chen1} linear MSRD codes of various square matrix sizes of the arbitrary length over a fixed field ${\bf F}_q$ were constructed.\\

We refer to \cite{MP191,NSZ21} for one-weight sum-rank codes.  Anticode bound for sum-rank codes and optimal such codes were given in \cite{CGLGMP}. Generalized sum-rank weights were introduced and their properties were studied in \cite{MP192}. Cyclic skew cyclic sum-rank codes and sum-rank BCH codes of the matrix size $n_1=\cdots=n_t=n$, $m_1=\cdots=m_t=m$ were proposed and studied in \cite{MP21} by the deep algebraic method. We refer to \cite{ALNWZ} for Hartmann-Tzeng bound and Roos bound on cyclic skew cyclic codes. Eigenvalue bounds for sum-rank codes were given in \cite{AKR} and these bounds were used to prove the nonexistence of MSRD codes for some parameters. It was proved in \cite{OPB} that random linear sum-rank codes attain the Gilbert-Varshamov-like bound with a high probability. In our previous papers \cite{Chen1}, larger linear sum-rank codes than these sum-rank BCH codes of the same minimum sum-rank distances were constructed. Welch-Berlemkamp decoding algorithm for linearized Reed-Solomon codes and sum-rank BCH codes in \cite{MP1,MP21}. Generic decoder for sum-rank codes was given in \cite{Puchinger} and the list-decoding of linearized Reed-Solomon codes was studied in \cite{PR}.  A general reduction of decoding in the sum-rank metric to the decoding in the Hamming metric was given in our paper \cite{Chen2}. For recent works on codes in the sum-rank metric, we refer to papers \cite{Berardini,Borello}.\\

{\bf Definition 1.2.} {\em Let ${\bf C} \subset {\bf F}_q^{(n_1,m_1), \ldots, (n_t, m_t)}$ be a code in the sum-rank metric, its covering radius $R_{sr}({\bf C})$ is the minimum radius $R_{sr}$ such that the balls $$B_{sr}({\bf x}, R_{sr})=\{{\bf y} \in {\bf F}_q^{(n_1,m_1), \ldots, (n_t, m_t)}: wt_{sr}({\bf y}-{\bf x}) \leq R_{sr}\},$$ centered at all codewords ${\bf x} \in {\bf C}$,  cover the whole space ${\bf F}_q^{(n_1,m_1), \ldots, (n_t, m_t)}$}. \\

The target of this paper is as follow. Given a positive integer $R<mt$, to find a small code in the sum-rank metric space ${\bf F}_q^{(m,m), \ldots, (m,m)}$ ($t$ blocks), such that its covering radius in the sum-rank metric is at most $R$.\\

We define the block length function $l_{q,m}(r, R)$ for linear covering codes in the sum-rank metric space ${\bf F}_q^{(m,m), \ldots, (m,m)}$ (with $t$ blocks). It is the smallest block length $t$, such that,  there is a block length $l_{q,m}(r,R)$ linear covering code in ${\bf F}_q^{(m,m), \ldots, (m,m)}$ (with $l_{q,m}(r,R)$ blocks) with the radius $R$ and the codimension at least $r$. This is a direct generalization of length functions of covering codes in the Hamming metric, which was introduced and studied in \cite{BPW}. From our construction and previous results in \cite{DMP,DMP2}, we can give some upper bounds on block length functions.\\

{\bf Definition 1.3.} {\em A code ${\bf C} \subset {\bf F}_q^{(n_1,m_1), \ldots, (n_t, m_t)}$ is called $(d, L)$ list-decodable in the sum-rank metric, if each ball $B_{sr}({\bf x}, d) \subset {\bf F}_q^{(n_1,m_1), \ldots, (n_t,m_t)}$,  contains at most $L$ codewords of ${\bf C}$ for each ${\bf x} \in {\bf F}_q^{(n_1,m_1), \ldots, (n_t, m_t)}$}.\\

For a family of sum-rank codes with block lengths going to infinity, if the list size $L$ is upper bounded by a fixed polynomial function of the block length, we say that this family of sum-rank codes is list-decodable. If the list size is exponential function of the block length, this family of codes is not list-decodable. One central topic about list-decodable codes in various finite metric space is to find the bound on list-decodability, see \cite{GRS}. For the asymptotic bound on list-decodability of rank-metric codes, we refer to \cite{WZ}.\\

Covering codes and list-decodable codes in the sum-rank metric degenerate to covering and list-decodable codes in the Hamming metric, if all matrix sizes are $1 \times 1$. Covering codes in the rank-metric were studied and some upper bounds and lower bounds for covering codes in the rank-metric were given in \cite{GY2,GY3}. The covering radius of the above Gabidulin code of the minimum rank-distance $n-k+1$ is $n-k$, see \cite{GY2}. In \cite{OLW,OLW1} some upper and lower bounds on covering codes in the sum-rank metric were given. However many expressions of lower and upper bounds in \cite{OLW, OLW1} are closely related to the volume of balls in the sum-rank metric and not given explicitly. List-decoding and some bounds on list-decodability of linearized Reed-Solomon codes was studied in \cite{PR}. However no upper bound on general list-decodable codes in the sum-rank metric has been given.\\

\subsection{Our Contributions}

In this paper, we study the covering problem in the sum-rank metric space ${\bf F}_q^{(m,m), \ldots, (m,m)}$. Block length functions of covering codes in the sum-rank metric are introduced, as the generalization of the length function introduced in \cite{BPW} for covering codes in the Hamming metric.  Some covering codes in the sum-rank metric via covering codes in the Hamming metric are constructed.  An upper bound on covering radius of some linear sum-rank codes constructed in \cite{Chen1} is presented. Then we give a general upper bound on sizes of covering sum-rank codes of the matrix size $m \times m$ and an upper bound on block length functions. \\

The space ${\bf F}_q^{(m,m), \ldots, (m,m)}$ is always the block length $t$ matrix space of the matrix size $m \times m$, consisting of all elements of the form $({\bf x}_1, \ldots, {\bf x}_t)$, where ${\bf x}_i$, $i=1, \ldots, t$, is a $m \times m$ matrix over ${\bf F}_q$. Bounds on list-decodable codes in the Hamming metric and the rank-sum metric were given in \cite{GRS,WZ}. The following bound on the list-decodablity of codes in the sum-rank metric is a natural generalization of bounds on list-decodable codes in the Hamming metric and rank-metric. This is the first such bounds for general list-decodable codes in the sum-rank metric.  we refer to \cite{PR} for bounds on list-decodablity of linearized Reed-Solomon codes.\\

{\bf Theorem 1.1.} {\em Let $m$ be a fixed positive integer and $\rho$ be a fixed real number satisfying $0<\rho<1$. Suppose that ${\bf C}_i$, $i=1,2, \ldots,$ is a family of codes in the sum-rank metric space ${\bf F}^{(m,m), \ldots, (m,m)}$ ($t_i$ blocks), with block lengths $t_i$ going to the infinity, and the rate satisfying
$$R=\lim_{i \longrightarrow \infty} R({\bf C}_i)=\frac{log_q |{\bf C}_i|}{m^2t_i} \geq 1-H_{q^m}(\rho)+\epsilon,$$  where $\epsilon$ is a fixed arbitrary small positive real number, then this family of codes is not $(\rho mt_i, L)$-list decodable in the sum-rank metric.}\\

The following bound is much stronger than the Singleton-like bound. From the Singleton-like bound, the sum-rank code in Corollary 1.1 has at most $2^{2(2t-d_{sr}+1)}$ codewords. When $n$ is large, the following strong Singleton-like bound is much stronger than the Singleton-like bound for sum-rank codes.\\

{\bf Corollary 1.1.} {\em 1) Let $t$ be the block length satisfying $t \geq 2^n-1$ and $d_{sr}$ is the minimum sum-rank distance satisfying $d_{sr}=8e+i$, $i=1,2, \ldots,7$, where $e$ is a positive integer satisfying $(2e-1)^{4e+2} \leq 2^n$. Then a binary sum-rank code of the block length $t$, the matrix size $2 \times 2$ and the minimum sum-rank distance $d_{sr}$ has at most $2^{2(2t-2n \lfloor \frac{d_{sr}}{8}\rfloor)}$ codewords.\\
2) When $d_{sr}=8e$, where $e$ is a positive integer satisfying $(2e-1)^{4e+2} \leq 2^n$, then a binary sum-rank code of the block length $t$, the matrix size $2 \times 2$ and the minimum sum-rank distance $d_{sr}$ has at most $2^{2(2t-2n \lfloor \frac{d_{sr}-1}{8}\rfloor)}$ codewords.}\\

Interestingly, the above strong Singleton-like bound can be compared with eigenvalue bounds for sum-rank codes developed in \cite{AKR}. Sum-rank-metric graphs were introduced in \cite{AKR} and eigenvalues of these graphs are used to give upper bounds on ${\bf A}_q({\bf n}, {\bf m}, d)$. These explicit bounds given in Section V of \cite{AKR} are mainly applied to various matrix sizes sum-rank codes. \\

The third contribution of this paper is the following upper bound on lengths of general binary MSRD codes.\\

{\bf Corollary 1.2.} {\em Let ${\bf C}$ be a general binary MSRD code with the matrix size $2 \times 2$ and the minimum sum-rank distance $d_{sr}=8e+i$, $i=1,2,\ldots, 8$. Let $n$ be the smallest positive integer satisfying $(2e-1)^{4e+2} \leq 2^n$ and $n\geq 5$. Then the block length of ${\bf C}$ cannot be larger than $2^n-1$.}\\

To the best of knowledge, this is the first upper bound on block lengths of general bianry MSRD codes with the fixed square matrix size $2 \times 2$. We also refer to \cite[Example VI.9]{BGR} for and Section VI of \cite{AKR} for comparison.\\

\section{Construction of covering codes in the sum-rank metric}

In this section, we give our construction of covering codes in the sum-rank metric from covering codes in the Hamming metric.\\

\subsection{General construction}

The matrix size is restricted to the case $n_1=\cdots=n_t=m=m_1=\cdots=m_t$. For each ${\bf a} \in {\bf F}_{q^m}$, we denote $M_i({\bf a})$ the $m \times m$ matrix with the only nonzero $i$-th row ${\bf a}$. The key point of our construction of covering sum-rank metric codes is the identification $${\bf F}_q^{(m,m),\ldots, (m,m)}=\{M_0({\bf a}_0)+M_1({\bf a}_1)+\cdots+M_{m-1}({\bf a}_{m-1}): {\bf a}_0, \ldots: {\bf a}_{m-1} \in {\bf F}_{q^m}\}.$$ This is the isomorphism of the linear space over ${\bf F}_q$. It is clear that if the $m \times m$ matrix is of the form $M_{i_1}({\bf a}_1)+\cdots+M({\bf a}_h)$, than the rank is at most $h$.\\

Let ${\bf C}_0, \ldots, {\bf C}_{m-1}$ be $m$ general Hamming metric codes in ${\bf F}_{q^m}^t$ with the covering radius $R_0, \ldots, R_{m-1}$. The covering sum-rank metric code is defined by $SR_{covering}({\bf C}_0, \ldots, {\bf C}_{m-1})=\{M_0({\bf c}_0) +M_1({\bf c}_1) +\cdots+M_{m-1}({\bf c}_{m-1}): {\bf c}_0 \in {\bf C}_0, {\bf c}_1 \in {\bf C}_1, \ldots {\bf c}_{m-1} \in {\bf C}_{m-1}\}.$\\

{\bf Theorem 2.1.} {\em For each vector ${\bf x} \in {\bf F}_q^{(m,m), \ldots, (m,m)}$, there is a codeword ${\bf y}$ in the sum-rank code $SR_{covering}({\bf C}_0, \ldots, {\bf C}_{m-1})$, such that, $$d_{sr}({\bf x}, {\bf y}) \leq R_0+\cdots+R_{m-1}.$$ The size of the above covering code in the sum-rank metric is exactly $$|{\bf C}_0| \cdots |{\bf C}_{m-1}|.$$}\\

{\bf Proof.} We express ${\bf x} \in {\bf F}_q^{(m,m), \ldots, (m,m)}$ as $M_0({\bf x}_0)+M_1({\bf x}_1)+\cdots+M_{m-1}{\bf x}_{m-1})$, where ${\bf x}_i \in {\bf F}_{q^m}^t$. Then there exist codewords ${\bf c}_0 \in {\bf C}_0, \ldots, {\bf c}_{m-1}\in {\bf C}_{m-1}$ in such that $wt_H({\bf x}_i-{\bf c}_i) \leq R_i$, $i=0, \ldots, m-1$. Set ${\bf y}=M_0({\bf c}_0)+M_1({\bf c}_1)+\cdots+M_{m-1}({\bf c}_{m-1})$, $wt_{sr}({\bf x}-{\bf y})=wt_{sr}(M_0({\bf x}_0-{\bf c}_0)+\cdots+M_{m-1}({\bf x}_{m-1}-{\bf c}_{m-1}))$.\\

Notice that ${\bf x}_i-{\bf c}_i \in {\bf F}_{q^m}^t$, and has its Hamming weight at most $R_i$, $i=0, 1, \ldots, m-1$. Set $S_i=supp({\bf x}_i-{\bf c}_i)$, for $i=0,1, \ldots, m-1$. When a block position $j \in \{1, \ldots, t\}$, is in the intersection of $h$ such supports, the rank contributing to $wt_{sr}({\bf x}-{\bf y})=wt_{sr}(M_0({\bf x}_0-{\bf c}_0)+\cdots+M_{m-1}({\bf x}_{m-1}-{\bf c}_{m-1}))$ is at most $h$. Hence we count each block position in each such support one time, we have $wt_{sr}({\bf x}-{\bf y})=wt_{sr}(M_0({\bf x}_0-{\bf c}_0)+\cdots+M_{m-1}({\bf x}_{m-1}-{\bf c}_{m-1})) \leq R_0+\cdots+R_{m-1}$.\\

The following result follows from Theorem 2.1 immediately.\\

{\bf Corollary 2.1.} {\em Let ${\bf C}$ be a Hamming metric codes in ${\bf F}_{q^m}^t$ with the covering radius $R$. Then $SR_{covering}({\bf C})=SR_{covering}({\bf C}, \ldots, {\bf C})$ is a covering code in the sum-rank metric with the radius at most $mR$. The size of this covering code in the sum-rank metric is $|{\bf C}|^m$.}\\

{\em Example 2.1.} Let $q$ be a prime power, $m$ and $n$ be two fixed positive integers. Then from the main result \cite{Le}, the covering radius of the first order $q^m$-ary Reed-Muller code $RM_{q^m}(1, n)$ is at most $$(q^m-1)q^{m(n-1)}-q^{m(\frac{n}{2}-1)}.$$ The length of the code is $q^{mn}$ and the code has $q^{m(n+1)}$ codewords. From Theorem 2.1, we have a covering code in the sum-rank metric space ${\bf F}_q^{(m,m), \ldots, (m,m)}$ ($q^{mn}$ blocks) with the radius at most $$m((q^m-1)q^{m(n-1)}-q^{m(\frac{n}{2}-1)}).$$ The size of this covering code is $q^{m(n+1)}$.\\

Since there are lot of results on covering codes in the Hamming metric, see \cite{CHLL} and references therein, we can construct many good covering codes in the sum-rank metric, based on the above construction. For example, from the redundancy bound, we can construct a covering code in the sum-rank metric with the radius $R_{sr}$ and the size at most $q^{m^2t-mR_{sr}}$. This bound was also obtained in \cite{OLW}. On the other hand, The Delsarte bound, see \cite[Theorem 11.3.3]{HP} or \cite{Delsarte}, asserts that if the dual code ${\bf C}^{\perp}$ has at most $h$ nonzero weights, then the covering radius of ${\bf C}$ is at most $h$.  Corollary 2.2 follows from Theorem 2.1 and the Delsarte theorem immediately.\\

{\bf Corollary 2.2.} {\em Let ${\bf C}^{\perp} \subset {\bf F}_{q^m}^t$ be a linear code with $h$ nonzero weights. Then we have a covering code $SR_{covering}({\bf C}) \subset {\bf F}_q^{(m,m), \ldots, (m,m)}$ in the sum-rank metric, with the size $|{\bf C}|^m$ and the covering radius at most $mh$.}\\

{\em  Example 2.2.} The Kasami codes proposed in  \cite{Kasami} are binary linear $[2^{2m}-1, 3m, 2^{2m-1}-2^{m-1}]_2$ codes with three nonzero weights, $w_1=2^{2m-1}-2^{m-1}$, $w_2=2^{2m-1}$ and $w_3=2^{2m-1}+2^{m-1}$, $m=1, 2, \ldots$. Its dual is a binary linear $[2^{2m}-1, 2^{2m}-1-3m]_2$ code ${\bf C}$ with the covering radius at most $3$.
${\bf C} \otimes {\bf F}_{2^n} \subset {\bf F}_{2^n}^{2^{2m}-1}$ is a covering code with the radius $3n$. Then we have a covering code in the sum-rank metric space ${\bf F}_2^{(n,n), \ldots, (n,n)}$ with the radius $3n^2$ and the cardinality $2^{n^2(2^{2m}-1-3m)}$.\\

{\em Example 2.3.} Let $m$ be a positive integer satisfying $m \equiv 0$ $mod$ $6$. Then a binary cyclic $[2^m-1, \frac{5m}{2}]_2$ code with $7$ nonzero weights was constructed in \cite{WZD}. From the Delsarte upper bound, the dual code is a binary linear $[2^m-1, 2^m-1-\frac{5m}{2}]_2$ code with the covering radius at most $7$. ${\bf C} \otimes {\bf F}_{2^n} \subset {\bf F}_{2^n}^{2^m-1}$ is a covering code with the radius $7n$.  Then we have a covering code in the sum-rank metric space ${\bf F}_2^{(n,n), \ldots, (n,n)}$ with the radius $7n^2$ and the cardinality $2^{n^2(2^m-1-\frac{5m}{2})}$.\\

Let ${\bf C}_i \subset {\bf F}_{q^m}^t$ be a covering code in the Hamming metric with the radius $R_i$, $i=0, 2, \ldots, m-1$. Set $SR({\bf C}_0, \ldots, {\bf C}_{m-1})=\{{\bf c}_0 x+{\bf c}_1 x^q+\cdots+{\bf c}_{m-1} x^{q^{m-1}}: {\bf c}_0 \in {\bf C}_0, {\bf c}_1 \in {\bf C}_1, \ldots {\bf c}_{m-1} \in {\bf C}_{m-1}\}.$\\

{\bf Theorem 2.2.} {\em The covering radius of the above covering code $SR({\bf C}_0, \ldots, \\{\bf C}_{m-1})$ in the sum-rank metric is at most $m(R_0+R_1+\cdots+R_{m-1})$.}\\

{\bf Proof.} Each vector ${\bf v}$ in ${\bf F}_q^{(m,m), \ldots, (m,m)}$ can be expressed as ${\bf v}_0x+\cdots+{\bf v}_{m-1}x^{q^{m-1}}$, where ${\bf v}_i \in {\bf F}_{q^m}^t$, $i=0, 1, \ldots, m-1$. We can find codewords ${\bf c}_0 \in {\bf C}_0, \ldots, {\bf c}_{m-1} \in {\bf C}_{m-1}$, such that, $wt_H({\bf v}_i-{\bf c}_i) \leq R_i$, $i=0, \ldots, m-1$. Set ${\bf c}={\bf c}_0x+\cdots+{\bf c}_{m-1}x^{q^{m-1}}$, then $wt_{sr}({\bf v}-{\bf c}) \leq m(R_0+\cdots+R_{m-1})$, since this vector are supported at most $R_0+\cdots+R_{m-1}$ positions.\\

\subsection{An asymptotical bound}

Let $m$ be a fixed positive integer, $t$ be a given positive integer and $q$ be a fixed prime power, for a given positive integer $R<n$, we denote $K_{q,m}(t, R)$ the minimal size of a code ${\bf C} \subset {\bf F}_q^{(m,m), \ldots, (m,m)}$ ($t$ blocks) with the covering radius in the sum-rank metric smaller than or equal to $R$. Set $ \frac{log_q K_{q, m}(t, \rho mt)}{m^2t}=k_t(q, \rho, m)$. Then we have the following upper bound.\\

{\bf Corollary 2.3.} {\em $$K_{q,m}(t, R) \leq K_{q^m}(t, \frac{R}{m})^m,$$ and $$k_t(q, \rho, m) \leq 1- H_{q^m}(\rho)+O(\frac{logt}{t}).$$}\\

{\bf Proof.} The first follows from Theorem 2.1 and 2.2. We have $\frac{log_q K_{q,m}(t, \rho mt)}{m^2t} \leq \frac{m^2log_{q^m}K_{q^m}(t, \rho t)}{m^2t}=\frac{log_{q^m} K_{q^m}(t, \rho t)}{t}$. The second conclusion follows immediately.\\

\subsection{Block length functions}

The following simple fact about block length functions is similar to length functions of covering codes in the Hamming metric, see \cite{BPW} and \cite[Chapter 11]{HP}.\\

{\bf Proposition 2.1.} {\em For each block length $t \geq l_{q,m}(r, R)$, there is a covering code ${\bf C} \subset {\bf F}_q^{(m,m), \ldots, (m,m)}$ ($t$ blocks) with the radius at most $R$ and the codimension at least $r$.}\\

{\bf Proof.} Let ${\bf C}_1$ be a block length $t$ covering sum-rank code with the radius at most $R$ and codimension $r$. Then we construct the block length $t\geq l_{q,m}(r, R)$ covering sum-rank code, $${\bf C}={\bf C}_1 \times ({\bf F}_q^{(m,m)})^{t-l_{q,m}(r,R)}.$$ It is easy to verify that the covering radius and the codimension of ${\bf C}$ are the same as the radius and the codimension of the covering radius of ${\bf C}_1$.\\

From Theorem 2.1, we have the following upper bound on the block length function of covering codes in the sum-rank metric space ${\bf F}_q^{(m,m), \ldots, (m,m)}$.\\

{\bf Corollary 2.4.} {\em $$l_{q,m}(r, R) \leq l_{q^m}(\frac{r}{m}, \frac{R}{m}).$$}\\

Then from the main result in \cite{DMP2} we have the following result.\\

{\bf Theorem 2.4.} {\em Let $u$ be an arbitrary positive integer. Then $$l_{q,m}(uR+m,R) \leq c q^{m\frac{(u-1)R+m}{R}} (mlnq)^{\frac{m}{R}}$$.}\\

{\em Example 2.4.} Let $q_1$ be an odd prime power and $n$ be an odd positive integer. Then a linear $[q_1^n, 2n+1]_{q_1}$ code with $4$ nonzero weights was constructed in Theorem 2 of \cite{LLQ}. The dual code is a linear $[q_1^n, q_1^n-1-2n]_{q_1}$ code with the covering radius at most $4$, from the Delsarte theorem \cite[Theorem 11.3.3]{HP}. Then $l_{q_1}(2n+1, 4) \leq q_1^n$.\\

When $q_1=q^m$, where $m$ is a fixed positive integer, $q$ is an odd prime power. We have the following upper bound on the block length function $l_{q,m}(m^2(2n+1),4m^2) \leq q^{nm}$.\\

\subsection{Covering sum-rank codes of the matrix size $2 \times 2$}

Let $K_4(t, R)$ be the smallest size of quaternary covering codes in ${\bf F}_4^t$ with the radius at most $R$. Then in \cite[Table 6.3]{CHLL}, some upper bounds on $K_4(t,R)$ for $t \leq 10, R\leq 6$ were given. Then we have the following results on $K_{2,2}(t, R)\leq K_4(t, \frac{R}{2})^2$. $K_{2,2}(10, 2) \leq 49152^2$, $K_{2,2}(10, 4) \leq 4096^2$, $K_{2,2}(10, 6) \leq 1024^2$, $K_{2,2}(10,8) \leq 208^2$ and $K_{2,2}(10, 10) \leq 2^{12}$. The resulted covering codes in the sum-rank metric are obtained from one covering code in the Hamming metric.\\

However, we can take different two covering codes in the Hamming metric, with radius $R_1$ and $R_2$ satisfying $$R_1+R_2=R_{sr}.$$ For example, from the upper bound $K_4(10,2) \leq 4096$ and $K_4(10,4) \leq 208$, a covering code in the sum-metric with $4096 \cdot 208$ the radius $6$ is obtained. Then we have $K_{2,2}(10, 6) \leq 2^{16} \cdot 13$. The is better than the above construction using one covering code in the Hamming metric with the radius $3$. Similarly, we have $K_{2,2}(10,8) \leq 2^{16}$, which is better than $K_{2,2}(10,8) \leq 208^2$, $K_{2,2}(10,10) \leq 2^8 \cdot 13$, which is better than $K_{2,2}(10,10) \leq 2^{12}$.\\

{\em Example 2.5.} $K_{2,2}(10,6) \leq 2^{16} \cdot 13$ means that there is a size $2^{16} \cdot 13$ covering code ${\bf C}$ in the sum-rank metric, of the block length $10$, the matrix $2 \times 2$. The covering radius of this code is $6$. That is, in the sum-rank metric space ${\bf F}_2^{(2,2), \ldots, (2,2)}$ ($10$ blocks), we can arrange $2^{16} \cdot 13$ codewords, such that, for any vector ${\bf x} \in {\bf F}_2^{(2,2), \ldots, (2,2)}$ ($10$ blocks), there is a codeword ${\bf y} \in {\bf C}$ satisfying $wt_{sr}({\bf x}-{\bf y}) \leq 6$.\\

From \cite[Table 6.3]{CHLL}, we give the following table of upper bounds on $K_{2,2}(t,R)$ of covering codes in the sum-rank metric.\\

\begin{longtable}{|l|l|l|l|l|l|l|l|}
\caption{\label{tab:A-q-5-3} Upper bounds on $K_{2,2}(t, R)$.}\\ \hline
$t, R$&2&4&6&8&10&12 \\ \hline
$6$& $65536$  & $2704$ & $208$&$16$&$4$&$1$ \\ \hline
$7$ & $1048576$  &$16384$ & $1024$&$128$&$16$&$16$ \\ \hline
$8$ &$11943936$  &$147456$ & $9216$&$768$&$64$&$16$\\ \hline
$9$ &$150994944$ & $1048576$ & $65536$&$4096$&$256$&$16$\\ \hline
$10$ &$2415919104$  & $16777216$ & $851968$&$65536$&$3328$&$256$ \\ \hline
\end{longtable}

\subsection{Covering radii of some explicit linear sum-rank codes}

Many larger explicit sum-rank codes of the form $SR({\bf C}_0, \ldots, {\bf C}_{m-1})$ then sum-rank BCH codes in \cite{MP21} with the same minimum sum-rank distances have been obtained in our previous paper \cite{Chen1}. From the above Theorem 2.2, an upper bound on their covering radius in the sum-rank metric can be given as in the following example.\\

{\em Example 2.6.} We consider the binary primitive BCH $[2^n-1, 2^n-1-ne, 2e+1]_2$ code $BCH(e,n)$, see \cite[Page 262]{CHLL}. It can be considered as quaternary code $BCH(e,n)_4$ with the covering radius at most $4e$, when $(2e-1)^{4e+1}\leq 2^n$, see \cite[Theorem 10.3.1]{CHLL}.\\

Let $e$ be an even positive integer satisfying $(2e-1)^{4e+1}\leq 2^n$. From the construction in \cite{Chen1}, $SR(BCH(e,n)_4, BCH(\frac{e}{2},n)_4)$ is a block size $2^n-1$, matrix size $2 \times 2$, dimension $4(2^n-1)-3ne$ sum-rank code with the minimum sum-rank distance at least $2e+1$. From Theorem 2.3, the covering radius of $SR(BCH(e,n)_4, BCH(\frac{e}{2},n)_4)$ in the sum-rank metric is at most $2(4e+2e)=12e$. When $e=4$, this is a block length $2^n-1$, dimension $4(2^n-1)-12n$ sum-rank code with the minimum sum-rank distance at least $9$. The covering radius of this code in the sum-rank metric is at most $48$.\\

\section{An asymptotic bound on list-decodability of sum-rank codes}

We have the following upper bound on sizes of list-decodable codes in the sum-rank metric, with an auxiliary covering code.\\

{\bf Theorem 3.1} {\em Let ${\bf C}_1 \subset {\bf F}_q^{(m,m), \ldots, (m,m)}$ ($t$ blocks) be a covering code in the sum-rank metric with the radius $d$. Let ${\bf C} \subset {\bf F}_q^{(m,m), \ldots, (m,m)}$ be an $(d, L)$-list decodable sum-rank metric code. Then we have the following upper bound on the size of ${\bf C}$, $$|{\bf C}| \leq L \cdot |{\bf C}_1|.$$}\\

{\bf Proof.} In each ball centered at codewords of ${\bf C}_1$ with the radius $d$, there are at most $L$ codewords of ${\bf C}$. Since these balls cover the whole space ${\bf F}_q^{(m,m), \ldots, (m,m)}$ ($t$ blocks), there are at most $L \cdot |{\bf C}_1|$ codewords in ${\bf C}$.\\

Then Corollary 3.1 and 3.2 follow from Theorem 3.1 and Theorem 2.1 directly. \\

{\bf Corollary 3.1.} {\em Let ${\bf C}_1 \subset {\bf F}_{q^m}^t$ be a Hamming metric covering code with the size $|{\bf C}_1|$ and the covering radius $R=\frac{d}{m}$. Let ${\bf C} \subset {\bf F}_q^{(m,m), \ldots, (m,m)}$ be an $(d, L)$-list decodable code in the sum-rank metric. Then we have the following upper bound on the size of ${\bf C}$, $$|{\bf C}| \leq L \cdot |{\bf C}_1|^m.$$}\\

{\bf Corollary 3.2.} {\em Let ${\bf C}_1 \subset {\bf F}_{q^m}^t$ be a Hamming metric covering code with the size $|{\bf C}_1|$ and the covering radius $R$. Then any sum-rank code ${\bf C}$ of the block length $t$, the matrix size $m \times m$, and the minimum sum-rank distance at least $2mR+1$, has at most $|{\bf C}_1|^m$ codewords.}\\

We have the following simple upper bound on sizes of $(d,L)$-list decodable codes in the sum-rank metric. The lower bound on list size $L$ can be obtained directly from the following bound. Then when code is too large, the following bound gives a bound on list-decodability of codes in the sum-rank metric.\\

{\bf Corollary 3.3.} {\em Let ${\bf C} \subset {\bf F}_q^{(m,m), \ldots, (m,m)}$ be a block length $t$ $(d, L)$-list decodable code. Then we have the following upper bound on the size of ${\bf C}$, $$|{\bf C}| \leq L \cdot q^{m^2t-md}.$$}\\

{\bf Proof.} We construction a covering code ${\bf C}_1\subset {\bf F}_q^{(m,m), \ldots, (m,m)}$ with the radius $d$ and at most $q^{m^2t-md}$ codewords as described after Corollary 2.1. The conclusion follows immediately.\\

From Corollary 2.3, we have the following asymptotic bound on the list-decodability of sum-rank codes.\\

{\bf Theorem 3.2.} {\em Let $m$ be a fixed positive integer and $\rho$ be a fixed real number satisfying $0<\rho<1$. Suppose that ${\bf C}_i$, $i=1,2, \ldots,$ is a family of codes in the sum-rank metric space ${\bf F}^{(m,m), \ldots, (m,m)}$ ($t_i$ blocks), with block lengths $t_i$ going to the infinity, and the rate satisfying
$$R=\lim_{i \longrightarrow \infty} R({\bf C}_i)=\frac{log_q |{\bf C}_i|}{m^2t_i} \geq 1-H_{q^m}(\rho)+\epsilon,$$ where $\epsilon$ is a fixed arbitrary small positive real number. Then this family of sum-rank code is not $(\rho mt_i, L)$-list decodable.}\\

{\bf Proof.} From Corollary 2.3, there is a covering code in ${\bf F}^{(m,m), \ldots, (m,m)}$ ($t_i$ blocks), with the  block length $t_i$, at most $q^{m^2(k_{t_i}(q,m,\rho mt_i)+\epsilon')}$ codewords, and the radius $\rho mt_i$. Here $\epsilon'$ is another arbitrary small positive real number. Then in the list-decodable sum-rank code ${\bf C}_i$, there are at most $L \cdot q^{m^2(k_{t_i}(q,m,\rho mt_i)+\epsilon')}$ codewords from Theorem 3.1. Then the rate satisfies $R({\bf C}_i) \leq k_{t_i}(q, m, \rho)+\epsilon'$ and the conclusion follows directly.\\

Notice that when $m=1$, this is just the asymptotic bound on list-decodability of codes in the Hamming metric, see \cite{GRS}.\\

\section{Strong Singleton-like bounds on general sum-rank codes}

Covering codes in Example 2.2 and 2.3 can be used to give the following two strong Singleton-like bounds for sum-rank codes with the minimum sum-rank distance $6n+1$ or $14n+1$ from Theorem 2.1 and 3.1.\\

{\bf Proposition 4.1.} {\em 1) Let $t$ be the block length satisfying $t \geq 2^{2m}-1$ and $d_{sr}=6n^2+1$ be the minimum sum-rank distance. Then a binary sum-rank code in ${\bf F}_2^{(n,n), \ldots, (n,n)}$ ($t$ blocks) with the minimum sum-rank distance $d_{sr}$ has at most $2^{n^2(t-3m)}$ codewords.\\
2) Let $m$ be a positive integer satisfying $m \equiv 0$ $mod$ $6$, $t$ be the block length satisfying $t \geq 2^m-1$ and $d_{sr}=14n^2+1$ be the minimum sum-rank distance. Then a binary sum-rank code in ${\bf F}_2^{(n,n), \ldots, (n,n)}$ ($t$ blocks) with the minimum sum-rank distance $d_{sr}$ has at most $2^{n^2(t-\frac{5m}{2})}$ codewords.}\\

Notice that from Singleton-like bound, the sum-rank code in 1) has at most $2^{n^2(t-6n)}$ codewords. It is clearer our strong Singleton-like bound is much better, when $t \geq 2^{2m}-1$ and $m>2n$.\\

Binary primitive BCH codes $BCH(e, n)$ of the length $2^n-1$ can be considered as code $BCH(e, n)_{2^m}=BCH(e,n) \otimes {\bf F}_{2^m}$ with the covering radius at most $2me$, see \cite[Theorem 10.3.1]{CHLL}. Then we have the following strong Singleton-like bound, following from Theorem 2.1 and Theorem 3.1.\\

{\em Theorem 4.1.} {\em 1) Let $t$ be the block length satisfying $t \geq 2^n-1$ and $d_{sr}$ be the minimum sum-rank distance satisfying $d_{sr}=2m^2e+i$, $i=1,2, \ldots,2m^2-1$, where $e$ is a positive integer satisfying $(2e-1)^{4e+2} \leq 2^n$. Then a binary sum-rank code in ${\bf F}_2^{(m,m), \ldots, (m,m)}$ ($t$ blocks) with the minimum sum-rank distance $d_{sr}$ has at most $2^{m^2(t-n \lfloor \frac{d_{sr}}{2m^2}\rfloor)}$ codewords.\\
2) When $d_{sr}=2m^2e$, where $e$ is a positive integer satisfying $(2e-1)^{4e+2} \leq 2^n$, then a binary sum-rank code in ${\bf F}_2^{(m,m), \ldots, (m,m)}$ ($t$ blocks) with the minimum sum-rank distance $d_{sr}$ has at most $2^{m^2(t-n \lfloor \frac{d_{sr}-1}{2m^2}\rfloor)}$ codewords.}\\

When block lengths are large, for example, $n\geq m^2+1$, and the $d_{sr}$ satisfies the condition $(2e-1)^{4e+1}<2^n$, the above strong Singleton-like bound is much stronger than Singleton-like bound for sum-rank codes. On the other hand, it is clear if smaller covering codes in the sum-rank metric can be constructed, the above bound can be improved.\\

When the minimum sum-rank distance $d_{sr}=2R+1$ is fixed, we have the following strong Singleton-like bound from the block length function.\\

{\em Theorem 4.2.} {\em Let $u$ be arbitrary positive integer. Suppose that ${\bf C} \subset {\bf F}_q^{(m,m), \ldots, (m,m)}$ ($t$ blocks) is a sum-rank code with the minimum sum-rank distance $d_{sr}=2R+1$. If $$t \geq c q^{m\frac{(u-1)R+m}{R}} (mlnq)^{\frac{m}{R}} \geq l_{q,m}(uR+m, R),$$ where $c$ is an universal constants independent of $q$ and $m$, then ${\bf C}$ has at most $q^{m^2t-m-u \cdot \frac{(d_{sr}-1)}{2}}$ codewords.}\\

{\bf Proof.} The conclusion follows from Theorem 3.1 for $(R,1)$-list-decodable codes, the definition of block length function for covering codes in the sum-rank metric and Theorem 2.4.\\

The above result asserts that, in dimension of the Singleton-like bound, we can have $-u \cdot \frac{d_{sr}-1}{2}$ for arbitrary large positive integer $u$, if the block length is sufficiently larger than a function of $u$.  This is a quite strong Singleton-like bound, when block lengths are sufficiently large.\\

From Example 2.4, we have the following strong Singleton-like bound for sum-rank codes.\\

{\bf Theorem 4.3.} {\em Let $q$ be an odd prime power, $n$ and $m$ be two fixed positive integers, and $t$ be a positive integer satisfying $t \geq q^{mn}$. Suppose that ${\bf C} \subset {\bf F}_q^{(m,m), \ldots, (m,m)}$ ($t$ blocks) is a sum-rank codes with the minimum sum-rank distance $8m^2+1$. Then ${\bf C}$ has at most $q^{m^2(t-2n-1)}$ codewords.}\\

Hence if $n\geq 4$, and the block length $t \geq q^{mn}$, the above strong Singleton-like bound is stronger than the general Singleton-like bound for sum-rank codes.\\

\section{Upper bounds on block lengths of general binary MSRD codes}

It is interesting and important to understand parameters for which there is no MSRD code. Several such results have been obtained in \cite{BGR,AKR}. From Proposition 4.1 we have the following result.\\

{\em Corollary 5.1.} {\em Let ${\bf C} \subset {\bf F}_2^{(n,n), \ldots, (n,n)}$ ($t$ blocks) be a general binary MSRD code with the minimum sum-rank distance $d_{sr}=6n^2+1$. Then the block length $t$ cannot be larger than $2^{4n}-1$.}\\

We have the following upper bound on binary MSRD codes from strong Singleton-like bounds proved in the previous section.\\

{\bf Corollary 5.2.} {\em Let ${\bf C} \subset {\bf F}_2^{(m,m), \ldots, (m,m)}$ ($t$ blocks) be a general binary MSRD code with the minimum sum-rank distance $d_{sr}=2m^2e+i$, $i=1,2,\ldots, 2m^2$. Let $n$ be the smallest positive integer satisfying $(2e-1)^{4e+2} \leq 2^n$ and $n\geq 2m^2+1$. Then the block length $t$ of ${\bf C}$ cannot be larger than $2^n-1$.}\\

There is no previous known general upper bound on lengths of binary MSRD codes in ${\bf C} \subset {\bf F}_2^{(m,m), \ldots, (m,m)}$ ($t$ blocks) in the literature. This is the first such upper bound. On the other hand, when matrix sizes of sum-rank codes are not the same, there is no such restriction on block lengths of MSRD codes, see \cite{MP20,Chen1}.\\

Similarly, we have the following upper bound on block lengths of $q$-ary MSRD codes, where $q$ is an odd prime power, from Theorem 4.3.\\

{\bf Corollary 5.3.} {\em Let $q$ be an odd prime power. Let ${\bf C} \subset {\bf F}_q^{(m,m), \ldots, (m,m)}$ ($t$ blocks) be a general $q$-ary MSRD code with the minimum sum-rank distance $d_{sr}=8m^2+1$. Then the block length $t$ of ${\bf C}$ cannot be larger than $q^{4m}$.}\\

\section{Conclusion}

Sum-rank metric coding is a generalization and combination of the Hamming metric coding and the rank-metric coding. In this paper, we constructed covering codes in the sum-rank metric from covering codes in the Hamming metric. Block length functions of covering codes in the sum-rank metric were introduced and studied. As applications, strong Singleton-like bounds, which are much stronger than the Singleton-like bound for sum-rank codes, were obtained. Then upper bounds on block lengths of general binary MSRD codes were presented. An asymptotic bound on list-decodability of codes in the sum-rank metric was proved.  A table of short binary covering codes in the sum-rank metric with the matrix size $2 \times 2$ and small radii was given. It seems interesting and necessary to construct smaller covering codes in the sum-rank metric. Then strong Singleton-like bound and upper bound on lengths of general binary MSRD codes can be improved.\\

\end{document}